\documentclass[a4letter,12pt]{article}
\usepackage{amsmath}
\usepackage{amsfonts}
\usepackage{latexsym}
\usepackage{graphics} 
\usepackage{graphicx,grffile}
\usepackage{subfig}
\usepackage{multicol}
\usepackage{times}
\usepackage{latexsym}
\usepackage{enumerate}
\usepackage{epsfig}
\usepackage{amssymb}
\usepackage{amsfonts}
\usepackage[english]{babel}
\usepackage{graphics}
\usepackage{placeins}
\usepackage{amsbsy}
\usepackage{setspace}
\usepackage{float}
\usepackage{authblk}
\usepackage{bbm}
\usepackage{color}
\usepackage[normalem]{ulem}
\usepackage{multirow}
\usepackage[hidelinks]{hyperref}
\usepackage{enumitem}
\usepackage{booktabs}
\usepackage{calrsfs}
\DeclareMathOperator*{\argmax}{arg\,max}
\DeclareMathAlphabet{\pazocal}{OMS}{zplm}{m}{n}
\usepackage[letterpaper, margin=1in]{geometry}
 
\usepackage[symbol*]{footmisc}

\DefineFNsymbolsTM{otherfnsymbols}{%
  \textasteriskcentered *
  \textsection   \mathsection
  \textbardbl    \|%
}%

\setfnsymbol{otherfnsymbols}

\newcommand{\blind}{0}

\begin{document}
\def\spacingset#1{\renewcommand{\baselinestretch}%
{#1}\small\normalsize} \spacingset{1}
\if0\blind
{
  \author[1]{Roberto Di Mari\thanks{roberto.dimari@unict.it}}
\author[2]{Roberto Rocci\thanks{roberto.rocci@uniroma2.it}}
\author[3]{Stefano Antonio Gattone\thanks{gattone@unich.it}}
\title{Constrained maximum likelihood estimation of clusterwise linear regression models with unknown number of components}
\affil[1]{Department of Economics and Business, University of Catania, Italy}
\affil[2]{Department of Economics and Finance, University of Rome Tor Vergata, Italy}
\affil[3]{Department of Philosophical and Social Sciences, Economics and Quantitative Methods, University G. d'Annunzio, Chieti-Pescara, Italy}
} 
\date{}
\maketitle

\fi

\if1\blind
{
  \bigskip
  \bigskip
  \bigskip
  \begin{center}
    {\LARGE\bf Constrained maximum likelihood estimation of clusterwise linear regression models with unknown number of components}
\end{center}
  \medskip
} \fi
%\newpage
\begin{abstract} 
We consider an equivariant approach imposing data-driven bounds for the variances to avoid singular and spurious solutions in maximum likelihood (ML) estimation of clusterwise linear regression models. We investigate its use in the choice of the number of components and we propose a computational shortcut, which significantly reduces the computational time needed to tune the bounds on the data. In the simulation study and the two real-data applications, we show that the proposed methods guarantee a reliable assessment of the number of components compared to standard unconstrained methods, together with accurate model parameters estimation and cluster recovery.    
\\

\end{abstract}

%\newpage
\spacingset{1.45}
\section{Introduction}\label{sec:intro}
In many applications within the various fields of social and physical sciences, investigating the relationship between a response variable and a set of explanatory variables is commonly of interest. Yet, the estimation of a single set of regression coefficients for all sample observations is often inadequate. To the purpose, finite mixture of conditional normal distributions can be used to estimate clusterwise regression parameters in a maximum likelihood context. Clusterwise linear regression is also known under the names of finite mixture of linear regressions or switching regressions (Alf\'o and Viviani, 2016; Quandt, 1972; Quandt and Ramsey, 1978; Kiefer, 1978). 

Let $\text{y}_1,\dots,\text{y}_n$ be a sample of independent observations drawn from the response random variable $\text{Y}_{i}$, each respectively observed conditionally on a vector of $J$ regressors $\mathbf{x}_{1},\dots,\mathbf{x}_{n}$. Let us assume $\text{Y}_{i} | \mathbf{x}_{i}$ to be distributed as a finite mixture of linear regression models, that is
\begin{equation}\label{eq:pop}
f(\text{y}_{i} | \mathbf{x}_{i} ; \pmb{\psi}) = {\sum_{g=1}^{G}p_{g}\phi_{g}(\text{y}_{i}|\mathbf{x}_{i},\sigma_{g}^{2},\pmb{\beta}_{g}}) = {\sum_{g=1}^{G}}p_{g}\frac{1}{\sqrt{2\pi\sigma_{g}^2}}\exp\bigg[-\frac{(\text{y}_{i}-\mathbf{x}_{i}^{\prime}\pmb{\beta}_{g})^2}{2\sigma_{g}^2}\bigg],
\end{equation}
%\begin{equation}
%\end{equation}
where $G$ is the total number of clusters and $p_{g},$ $\pmb{\beta}_{g},$ and $\sigma^2_{g}$ are respectively the mixing proportion, the vector of $J$ regression coefficients, and the variance term for the $g$-th cluster. The set of all model parameters to be estimated is given by $\pmb{\psi}=\{(p_{1},\dots,p_{G};\pmb{\beta}_{1},\dots,
\pmb{\beta}_{G};\sigma^{2}_{1},\dots,\sigma^{2}_{G}) \in \mathbb{R}^{G(J + 2) } : p_{1}+\dots+p_{G}=1, p_g > 0, \sigma_{g}^{2}>0$, for $g=1,\dots,G\}$.

The likelihood function can be formulated as
\begin{equation}\label{eq:Likeunc}
\mathcal{L}(\pmb{\psi}) = \prod_{i=1}^{n}\bigg\{{\sum_{g=1}^{G}}p_{g}{\frac{1}{\sqrt{2\pi{\sigma_{g}}^2}}}\exp\bigg[-\frac{(\text{y}_{i}-
\pmb{\text{x}}_{i}^{\prime}\pmb{\beta}_{g})^2}{2\sigma_{g}^2}\bigg]\bigg\},
\end{equation}
which is maximized in order to estimate $\pmb{\psi}.$ Alternatively to direct maximization, the EM algorithm (Dempster, Laird, and Rubin, 1977) is frequently used. 

Unlike finite mixtures of other densities, the parameters of a clusterwise linear regression model, under mild regularity conditions (Hennig, 2000) are identified. A well-known complication in ML estimation of mixtures of (conditional) normals with cluster-specific variances is that the likelihood function is unbounded (Kiefer and Wolfowitz, 1956; Day, 1969). This can be seen by noting that the likelihood function goes to infinity as one mixture's variance tends to zero and one of the sample observations has a zero residual on the corresponding component. Hence a global maximum does not exist. 

In spite of this, ML estimation does not fail. For switching regression with cluster-specific variances (heteroscedastic switching regressions), the likelihood equations have a consistent root (Kiefer, 1978). Yet, there is no obvious way of finding it when there is more than one local maximum. This has two practical consequences: EM degeneracy, and occurrence of spurious solutions. In the first case, the sequence of parameter estimates produced by the EM algorithm fails to converge because one or more cluster conditional variances go to zero. The second situation occurs instead when the algorithm converges to a non-meaningful local maximizer, typically characterized by an estimated mixture's component with a small number of points and a relatively small variance (McLachlan and Peel, 2000).

The problem of unboundedness has been tackled by a large number of authors and many different solutions have been proposed. A comprehensive review on the topic can be found in Garc\'ia-Escudero et al. (2017). See also Ritter (2014).

One strand of literature is based on the seminal work of Hathaway (1985) which, in order to have the likelihood function of univariate mixtures of normals bounded, suggested to impose a lower bound, say $c$, to the ratios of the scale parameters in the maximization step. The method is equivariant under linear affine transformations of the data. That is, if the data are linearly transformed, the estimated posterior probabilities do not change and the clustering remains unaltered.
In the context of switching regression, Philips (1991) and Xu, Tan and Zhang (2010) showed that if the true parameters satisfy the constraints, there is a global maximizer of the likelihood function which is consistent, asymptotically normal and efficient. Nevertheless, Hathaway's constraints are very difficult to apply within iterative procedures like the EM algorithm. In addition, how to properly choose $c,$ which controls the strength of the constraints, is an open issue.

Recently, in the multivariate case, Rocci et al., (2017) incorporated constraints on the eigenvalues of the component covariances of Gaussian mixtures that are tuned on the data (RGD method). Built upon Ingrassia (2004)'s reformulation, these constraints are an equivariant sufficient condition for Hathaway's constraints (in their multivariate extension) and do not require any prior knowledge of the mixture scale balance. Estimation is done in a familiar ML environment (Ingrassia and Rocci, 2007), and the data--driven selection of $c$ implements a cross-validation strategy. Di Mari et al., (2017) adapted the RGD method to clusterwise linear regression, further investigating its properties. 

Another possible approach for handling unboundedness is to modify the log-likelihood function by adding a penalty term, in which smaller values of the scale parameters are increasingly penalized. Representative examples can be found in Ciuperca et al. (2003), in the univariate case, and in Chen and Tan (2009), in the multivariate case.

A further alternative to constrained and penalized approaches is root selection, i.e. monitoring the local maximizers in order to discard nearly degenerate or spurious solutions  (McLachlan and Peel, 2000). However, this is not an easy task. Seo and Kim (2012) point out that a spurious solution is typically driven by a random localized pattern of a few observations in the data. Such observations are overfitted by one component of the mixture, heavily affecting the formation of the likelihood-based solutions. They suggest to take out such, say, $k$ observations with the highest likelihood (likelihood-based $k$-deleted method - Seo and Lindsay, 2010), or with the highest value for a score-based statistic (score-based $k$-deleted method), and select the solution with the highest $k$-deleted likelihood (or score-based statistic). Kim and Seo (2014) show that the score-based method in Seo and Kim (2012) can be well approximated by a gradient-based version of the $k$-deleted method, which is computationally more efficient.

The issue of unboundedness is not only an estimation problem, but makes one of the most complicated tasks in cluster analysis, i.e. selecting the number of components, more complicated. It is common to select the number of components by means of likelihood--based information criteria, like AIC or BIC. With degenerate and spurious solutions, unduly high likelihood values are very likely to result in distorted assessments of the number of clusters. 

The contribution of the present work is threefold. First, in line with recent research on multivariate mixtures of normals (Cerioli et al, 2017), we  investigate the use of the RGD method to \emph{regularize} the log-likelihood used for the Bayesian Information Criterion, which we evaluate at the constrained solution (Fraley and Raftery, 2007). Although this conjecture was already in Di Mari et al (2017)'s empirical examples, there have been no adequate test over a wide set of simulation scenarios yet. Second, we evaluate the model complexity in terms of estimated scales as a function of the tuning parameter, adapting Cerioli et al (2017)'s approach to the RGD constraints for clusterwise linear regression. The results from our simulation study demonstrate that model selection based on a \emph{regularized} likelihood, as well as a possible correction for the model complexity in terms of estimated scales, guarantees a reliable assessment of the number of mixture components. Nevertheless, this comes at the price of a higher computational cost - due to the way the tuning constant is chosen - compared to the unconstrained method. As third contribution, we present a computational shortcut to the RGD method for selecting $c$ based on the data, given $G$. This new and computationally faster version is based on the idea of the $k$-deleted method of Seo and Lindsay (2010) - used also in Seo and Kim (2012) and Kim and Seo (2014). In the simulation study and the two real-data applications we show that the proposed \emph{accelerated} method keeps up very well with the benchmark RGD method, in terms of model parameters estimation and cluster recovery. In addition, we show its soundness also in a model selection context.

The remainder of the paper is organized as follows. In Section \ref{sec:constrained}, we briefly review the constrained RGD method for clusterwise regression modeling and the cross--validation strategy to tune the constraints. Section \ref{sec:penBIC} describes how to carry out model selection with BIC based on the constrained estimator, and Section \ref{sec:modifiedkdel} introduces the computationally efficient alternative to the cross--validation strategy. The proposed methodologies are illustrated - and their performance evaluated - with a simulation study (Section \ref{sec:simulation}), and two real-data examples (Section \ref{sec:data}). Section \ref{sec:conclusion} concludes with a final discussion and some ideas for future research      

\section{The RGD method}\label{sec:constrained}
For univariate Gaussian mixtures, Hathaway (1985) proposed to maximize the log-likelihood under constraints of the kind
\begin{equation}\label{eq:hath}
 \min_{i \neq j} \frac{\sigma_{i}^{2}}{\sigma_{j}^{2}} \geq c \quad \text{with} \quad c \in (0,1]. 
\end{equation}
Hathaway's approach presents a strongly consistent global solution, no singularities, and a smaller number of spurious maxima. However, there is no easy way to implement the constraints into a feasible algorithm.

For ML estimation of the clusterwise linear regression model in Equation \eqref{eq:pop}, Di Mari et al. (2017) proposed relative constraints on the group conditional variances $\sigma^2_g$ of the kind

\begin{equation}\label{eq:constr1}
\sqrt{c} \leq \frac{\sigma^{2}_{g}}{\bar{\sigma}^{2}} \leq \frac{1}{\sqrt{c}},
\end{equation}

or equivalently
\begin{equation}\label{eq:constr2}
\bar{\sigma}^{2}\sqrt{c} \leq \sigma^{2}_{g} \leq \bar{\sigma}^{2}\frac{1}{\sqrt{c}}.
\end{equation}

The above constraints have the effect of shrinking the variances to a suitably chosen $\bar{\boldsymbol{\sigma}}^{2},$ the \emph{target} variance term, and the level of shrinkage is given by the value of $c.$ Easily implementable within the EM algorithm (Ingrassia, 2004; Ingrassia and Rocci, 2007), the constraints in~\eqref{eq:constr2} provide a sufficient condition for Hathaway's constraints - Equation ~\eqref{eq:hath} - to hold. This can be seen by noting that 
$$\frac{\sigma_{g}^{2}}{\sigma_{j}^{2}} = \frac{\sigma_{g}^{2}/\bar{\sigma}^{2}}{\sigma_{j}^{2}/\bar{\sigma}^{2}} \geq \frac{\sqrt{c}}{1/\sqrt{c}} = c$$

This type of constraints ensures the method to be equivariant (Di Mari et al., 2017), i.e. if the dependent variable is rescaled, and the \emph{target} variance $\bar{\boldsymbol{\sigma}}^{2}$ transformed accordingly, the linear predictor and the error's standard deviation are both on the new response scale. Perhaps most importantly for clustering, this leaves the estimates of the posterior probabilities unaltered, hence guaranteeing a final partition which does not depend on any previous data transformation or standardization.

A sensible choice of the tuning parameter $c$ is needed. Selecting $c$ jointly with the mixture parameters by maximizing the likelihood on the entire sample would trivially yield an overfitted scale balance approaching zero. Di Mari et al. (2017) proposed, for constrained estimation of clusterwise linear regression, a cross-validation strategy in order to let the data decide the optimal scale balance. The resulting scale balance can be seen as the most appropriate-to-the-data compromise between the heteroscedastic model (for $c\rightarrow 0,$ $\hat{\boldsymbol{\sigma}}^{2}_g$ equals the unconstrained ML estimate) and the homoscedastic model (when $c=1$, $\hat{\boldsymbol{\sigma}}^{2}_g = \bar{\boldsymbol{\sigma}}^{2})$. 
In particular, they consider partitioning $M$ times the data set $\{(\text{y}_i,\mathbf{x}_i)\}_{n}$ at random into a training set of size $n_S$ and a test set of size $n_{\bar{S}},$ where $n_S + n_{\bar{S}} = n.$ For the $m$-th partition, let $\widehat{\boldsymbol{\psi}} (c, S_m)$ be the constrained ML estimator based on the training set and $\ell_{\bar{S}_m} (\widehat{\boldsymbol{\psi}} (c, S_m))$ be the log-likelihood function evaluated at the test set. The cross-validated log-likelihood is
\begin{equation}\label{eq:cvlog}
\text{CV}(c) = \sum_{m=1}^{M} \ell_{\bar{S}_m} (\widehat{\boldsymbol{\psi}} (c, S_m)),
\end{equation}
which is the sum of the contribution of each test set to the log-likelihood. The optimal $c$ is found as the maximizer of the function in Equation \eqref{eq:cvlog}.  

The maximization of the cross-validated log-likelihood corresponds to the minimization of an unbiased estimator of the Kullback-Leibler divergence between the \emph{truth} and the model under consideration (Smyth, 1996; 2000). The logic behind its use is that it can be seen as function of $c$ only, and maximizing it handles the issue of overfitting as training and test sets are independent (Arlot and Celisse, 2000). The method has shown great promise in terms of quality of model parameters estimation; in the next Section, we propose its use for selecting the number of components. 

\section{Regularized BIC for model selection}
\label{sec:penBIC}

Likelihood--based information criteria, like the AIC and the BIC, are widely used to select the number of mixture components in probabilistic (model--based) clustering. Leroux (1992) showed that neither of the two underestimates the number of mixture components. Further studies showed that, whereby AIC tends to overestimate the number of components (Koehler \& Murphree, 1987), BIC consistently estimates it (Keribin, 2000). The BIC has two ingredients: the (negative) maximized mixture likelihood taking into account the overall fit of the model to the data, and a penalty term measuring model complexity and sample size.
Standard BIC has the form:
\begin{equation}\label{eq:bic}
\text{BIC} = -2 \log \mathcal{L}(\widehat{\pmb{\psi}})+ \eta \log(n),
\end{equation}
where $\eta = \underbrace{J+1}_{\text{regression coeff}} + \underbrace{G}_{\text{scales}} + \underbrace{G-1}_{\text{mixing proportions}}$ represents the number of free parameters to be estimated, and measures model complexity. It is self--evident that $\widehat{\pmb{\psi}}$ computed by using the unbounded likelihood could correspond to a degenerate or spurious solution, making BIC unreliable.

The constrained estimator eliminates degeneracy and reduces the number of spurious solutions (Hathaway, 1985), as the likelihood surface is regularized. How well the regularization is done depends on how the bounds are tuned: with an optimal data--driven selection strategy, we claim that the RGD approach can be used to compute the BIC for a sounder assessment of the number of components as the chance of overfitted solutions is greatly reduced. The BIC, computed at the constrained solution, is as follows:
\begin{equation}\label{eq:constrbic}
\text{BIC} = -2 \log \mathcal{L}(\widehat{\pmb{\psi}}_{cs})  + \eta \log(n).
\end{equation}

Similarly, to handle the unboundedness of the likelihood of multivariate Gaussian mixtures, Fraley and Raftery (2007) proposed to select the number of components by evaluating the BIC at the maximum a posteriori (MAP) estimator.

Notice that, in the BIC of Equation \eqref{eq:constrbic}, whatever the value of $c$, $\eta$ is fixed. In fact, different values of $c$ correspond to different  model complexity levels. Consider the case of $c$ close to $1$: the component variances are constrained to be similar to the target variance. In other words, much of their final estimated values comes from the target variance. In the opposite situation, a value of $c$ close to zero allows the component variances to (almost freely) vary. Based on similar considerations, Cerioli et al (2017)'s proposal amounts, in a clusterwise linear regression context, to measure the effective complexity due to the scales as the fraction $(1 - c) \times G$, yielding the following modified BIC

\begin{equation}\label{eq:bicCer}
\text{BIC}_{\text{mod}} = -2 \log \mathcal{L}(\widehat{\pmb{\psi}}_{cs})  + \eta^{*} \log(n),
\end{equation}
where $\eta^{*} = \underbrace{J+1}_{\text{regression coeff}} + \underbrace{(1-c)\times G}_{\text{free scales}} + \underbrace{G-1}_{\text{mixing proportions}}$.

In the simulation study and the empirical application, we will illustrate both model selection criteria of Equations \eqref{eq:constrbic} and \eqref{eq:bicCer} under different scenarios.

\section{A computationally efficient constrained approach}\label{sec:modifiedkdel}
In this Section, we first sketch the $k$-deleted method of Seo and Lindsay (2010), Seo and Kim (2012), and Kim and Seo (2014) in its na\"{i}ve formulation, i.e. the likelihood-based $k$-deleted method. Then, starting from their baseline idea, we propose a new, computationally faster, data driven method to tune $c.$  

\subsection{The likelihood-based $k$-deleted method}  

Singular or spurious solutions are characterized by one or a few observations having overly large log-likelihood terms compared to the rest of the sample. In such cases, these sample points end up dominating the overall log-likelihood. In order to identify such $k$ dominating observations, Seo and Kim (2012) suggested to use the individual log-likelihood terms, and then define the so called $k$-deleted log-likelihood as follows
\begin{equation}\label{eq:standardkdel}
\ell_{-k}(\pmb{\psi}) = \log \mathcal{L}(\pmb{\psi})  - \sum_{d=1}^k \log f(\text{y}_{(n-d+1)}; \pmb{\psi})
\end{equation}
where $f(\text{y}_{(1)}; \pmb{\psi})< \dots < f(\text{y}_{(n)}; \pmb{\psi})$ are the ordered values of the individual likelihood terms evaluated at $\pmb{\psi}$.  
Given the set of local maximizers $\boldsymbol{\Psi} = \{ \widehat{\pmb{\psi}}_{s}; \quad s=1,\dots,S \}$ previously found, the $k$-deleted log-likelihood is used as a criterion to select the root such that
\begin{equation}\label{eq:kdelcriterion}
\widehat{\pmb{\psi}}_{-k} = \argmax_{\boldsymbol{\psi} \in \boldsymbol{\Psi}} \{ \ell_{-k}(\pmb{\psi}) \}.
\end{equation}
In words, the very appealing feature of the (likelihood-based) $k$-deleted method is that it selects a solution among the ones already computed. The quantity in Equation \eqref{eq:standardkdel} represents how well the rest of the data are fitted after one removes the possible effect of overfitting a single or a few observations (Seo and Lindsay, 2010). On the other hand, whether effectively the method discards the spurious solutions in favor of the \emph{correct} one depends on actually computing the largest number of solutions possible. Exploring complicated likelihood surfaces will hence require well refined initialization strategies - possibly consisting of large sets of different starts.

\subsection{An efficient RGD approach}
 
For a given value of the tuning parameter $c,$ let $\widehat{\pmb{\psi}}_{cs}$ be the $s$-th local maximizer, with $s=1,\dots,S$, found maximizing (\ref{eq:Likeunc}) subject to (\ref{eq:constr2}). Let us define the $k$-deleted log likelihood as follows
\begin{equation}\label{eq:modifiedkdel}
\ell_{-k} (\widehat{\pmb{\psi}}_{cs}) = \log \mathcal{L}(\widehat{\pmb{\psi}}_{cs}) - \sum_{d=1}^k \log f(\text{y}_{(n-d+1)}; \widehat{\pmb{\psi}}_{cs}).
\end{equation}

The $k$-deleted log likelihood represents how well the rest of the data are fitted after one removes the possible effect of overfitting a single or a few observations. 

The constant $c$ is selected as follows
\begin{equation}\label{eq:modkdelcriterion}
c = \argmax_{0< c\leq 1} \{ \ell_{-k} (\widehat{\pmb{\psi}}_{cs}) \}.
\end{equation}

The negative term in Equation \eqref{eq:modifiedkdel} can be thought as a sort of penalty for spurious solutions. This term also eliminates the overfitting - in the same spirit as in Seo and Lindsay (2010), where it was used for selecting the bandwidth of their smoothed ML estimator. 

In addition, by implicitly selecting a maximizer for the constrained ML problem among constrained solutions, the method we propose does not depend on the initialization strategies employed as the number of spurious maximizers is already reduced in a constrained setup (Hathaway, 1985). Stating it differently, it applies a root selection approach - the $k$-deleted method - to a setup which already guarantees a smaller number of solutions to the ML problem.

\section{ML estimation}\label{sec:mlestimation}

Once a data--driven choice of $c$ is available, the RGD method requires a target variance as input. The most natural candidate, as argued in Di Mari et al. (2017), is the homoscedastic normal variance as, for $c=1,$ the RGD method's output is then simply the homoscedastic model. Notice that, if the target was chosen from another equivariant method, the RGD approach with scale balance equal to one would be estimating a common-variance model, with common variance equal to the target and the other model parameters estimated at their conditional ML values. 

ML parameter estimation, after the data--driven selection step, is done by means of Ingrassia and Rocci (2007)'s constrained EM (for details on the steps for clusterwise linear regression, see Di Mari et al, 2017).

\section{Numerical study}\label{sec:simulation} 
\subsection{Design}
The purpose of this simulation study is to address the following issues:
\begin{itemize}
\item how sensitive the efficient RGD approach is (ConK) to different choices of $k$;
\item how ConK compares with the standard RGD method (ConC), and with the homoscedastic (HomN) and heteroscedastic (HetN) models;
\item how the two reformulations of the BIC (Equations  \eqref{eq:constrbic} and \eqref{eq:bicCer}) perform under different scenarios, and how reliably they allow selecting the number of components compared to BIC computed at HomN and HetN solutions.
 \end{itemize}
Concerning the choice of $k$ (ConK approach), Seo and Kim (2012), for $P$-variate Gaussian mixtures, suggest choosing it between $P$ - where only one component is degenerate (or spurious) in all $P$ dimensions - and $P \times (G-1)$ - where $G-1$ components are degenerate in all $P$ dimensions. In the first part of the simulation study, we assess ConK in terms of accuracy of parameter estimates (average MSE of regression coefficients and component variances) and cluster recovery (adjusted \emph{Rand} index, Adj-Rand, of Hubert and Arabie, 1985)  for $k=\{1,2,J\times(G-1),n/10,n/5,n/2,n/1.25,n/1.11\}$, where $J$ is the number of regressors. We expect very similar results for the different $k$'s as the $k$-deleted method we implement in this paper is not a root selection method, but rather implemented to select a tuning parameter - similarly to what is done in Seo and Lindsay (2010) for bandwidth selection in their smoothed ML estimator. However note that we also test for values of $k$ very large compared to the sample size (e.g. $k = n/1.11$) to exclude the possibility that any $k \leq n$ works.  

In the second part of our simulation study, the performance of ConK  is compared with: 1) ConC, 2) the unconstrained algorithm with common (homoscedastic) component-scales (HomN), and 3) the unconstrained algorithm with different (heteroscedastic) component-scales (HetN). By taking the number of components as known, we evaluate the computation time, the accuracy of the model parameter estimates and of cluster recovery for all 4 methods. The target measures used for the comparisons are average MSE of the regression coefficients (averaged across regressors and groups) and of the component variances (averaged across groups) for estimation accuracy, and the adjusted \emph{Rand} index for cluster recovery.
  
In the third part of our simulation study, we take the number of components as unknown, and let each method select $G$ between 1 and $G^{*} + 2$  (where $G^{*}$ is the true number of groups) as the one for which the BIC (Equation \eqref{eq:bic}) is the lowest. For the constrained methods ConC and ConK, we do this exercise using the BIC reformulations of Equations \eqref{eq:constrbic} and \eqref{eq:bicCer}.

The data were generated from a clusterwise linear regression with 3 regressors and intercept, with 2, 3, and 4 components and sample sizes of 100 and 200. The class proportions considered were, respectively, $(0.5,0.5)^\prime$ and $(0.2,0.8)^\prime,$ and $(0.2,0.4,0.4)^\prime$ and $(0.2,0.3,0.5)^\prime,$ and $(0.25,0.25,0.25,0.25)^\prime$ and $(0.1,0.2,0.3,0.4)^\prime$. Regressors have been drawn from 3 independent standard normals, whereas regression coefficients have been drawn from U$(-1.5,1.5)$ and intercepts are $(4,9)^{\prime}$, $(4,9,16)^{\prime}$, and $(4,9,16,25)^{\prime}$ for the 2-component, 3-component and 4-component models respectively. The component variances have been drawn from Inv-Gamma$(3,1)$. 

For each of the 12 combinations sample size $\times$ mixing proportion, we generated 250 samples: for each sample and each algorithm - HomN, HetN, ConC, and ConK - we select the best solution (highest likelihood) out of 10 random starts.

\subsection{Results}

\FloatBarrier
\begin{table}
\centering
\resizebox{0.48\vsize}{!}{\begin{tabular}{lcccccccccc}
\hline \hline
$k$ && Avg MSE $\hat{\pmb{\beta}}$  && Avg MSE $\hat{\pmb{\sigma}}^{2}$ && $\text{Adj-Rand}$ && time && $c$ \\
 \hline

&&\multicolumn{9}{c}{Mixing proportions $(0.5,0.5)$}\\
     &&	       &&        &&	   &&        &&        \\     
 
$1$			 && 0.0135 && 0.0075 && 0.9619 && 0.1184 && 0.3024 \\
$2$ 			&& 0.0135 && 0.0074 && 0.9622 && 0.1239 && 0.3315 \\	
$J\times(G-1)$ && 0.0135 &&	0.0074 &&	0.9624 &&	0.1291 &&	0.3677 \\
$n/10$			 && 0.0135 &&	0.0073 &&	0.9625 &&	0.1679 &&	0.5360 \\
$n/5$ 			&& 0.0136 &&	0.0086 &&	0.9613 &&	0.1765 &&	0.6284 \\
$n/2$ 			&& 0.0137 &&	0.0110 &&	0.9601 &&	0.1837 &&	0.6526
 \\
$n/1.25$ 		&& 0.0134 &&	0.0094 &&	0.9620 &&	0.1723 &&	0.5589 \\
$n/1.11$ 		&& 0.0133 &&	0.0103 &&	0.9624 &&	0.1681 &&	0.5427 \\

\cmidrule{3-11} %

&&\multicolumn{9}{c}{Mixing proportions $(0.2,0.8)$}\\
     &&	       &&        &&	   &&        &&        \\
$1$ 			&& 0.0365 && 	0.0152 && 	0.9726 &&	0.1533 && 0.2353 \\
$2$ 			&& 0.0365 && 	0.0152 &&	0.9726 && 	0.1521 && 0.2365 \\
$J\times(G-1)$ 	&& 0.0365 &&	0.0152 &&	0.9726 && 	0.1490 &&	0.2355 \\
$n/10$ 			&& 0.0365 &&	0.0151 &&	0.9731 &&	0.1490 &&	0.2467 \\
$n/5$ 			&& 0.0364 &&	0.0152 &&	0.9729 &&	0.1534 &&	0.2456 \\
$n/2$ 			&& 0.0364 &&	0.0168 &&	0.9731 &&	0.1571 &&	0.2544 \\
$n/1.25$ 		&& 0.0367 &&	0.0154 &&	0.9726 &&	0.1910 &&	0.4322 \\
$n/1.11$ 		&& 0.0371 &&	0.0185 &&	0.9727 &&	0.2042 &&	0.5252 \\

\cmidrule{3-11}

&&\multicolumn{9}{c}{Mixing proportions $(0.2,0.4,0.4)$}\\
     &&	       &&        &&	   &&        &&        \\
$1$ 			&& 0.2009 && 0.0691 && 0.9517 && 0.5240 && 0.1201 \\
$2$ 			&& 0.2325 && 0.0916 && 0.9453 && 0.5431 && 0.1230 \\
$J\times(G-1)$  && 0.2247 && 0.0849 && 0.9455 && 0.5695 && 0.1667  \\
$n/10$ 			&& 0.2402 && 0.0850 && 0.9457 && 0.5870 && 0.1979  \\
$n/5$ 			&& 0.2084 && 0.0780 && 0.9465 && 0.7067 && 0.2430 \\
$n/2$ 			&& 0.1992 && 0.0872 && 0.9455 && 0.6443 && 0.3055 \\
$n/1.25$ 		&& 0.2005 && 0.1269 && 0.9474 && 0.6686 && 0.3301 \\
$n/1.11$ 		&& 0.2177 && 0.1582 && 0.9448 && 0.6791 && 0.3283 \\

\cmidrule{3-11}

&&\multicolumn{9}{c}{Mixing proportions $(0.2,0.3,0.5)$}\\
     &&	       &&        &&	   &&        &&        \\
$1$ 			&& 0.1256 && 0.0450 &&  0.9690 && 0.4710 && 0.0929 \\
$2$ 			&& 0.1254 && 0.0449 &&  0.9690 && 0.4775 && 0.1003 \\
$J\times(G-1)$  && 0.1248 && 0.0441 &&	0.9690 && 0.5079 &&	0.1226 \\
$n/10$ 			&& 0.1239 && 0.0427 &&	0.9696 && 0.5236 &&	0.1429 \\
$n/5$ 			&& 0.0934 && 0.0397 &&	0.9696 && 0.5413 &&	0.1820 \\
$n/2$ 			&& 0.0916 && 0.0459 &&	0.9699 && 0.5636 &&	0.2770 \\
$n/1.25$ 		&& 0.0882 && 0.0855 &&	0.9676 && 0.6018 &&	0.3314 \\
$n/1.11$ 		&& 0.1085 && 0.1213 &&	0.9655 && 0.5810 &&	0.3537 \\

\cmidrule{3-11}

&&\multicolumn{9}{c}{Mixing proportions $(0.25,0.25,0.25,0.25)$}\\
     &&	       &&        &&	   &&        &&        \\
$1$ 			&& 1.0741 && 0.2464 &&  0.9083 &&   1.5461 &&   0.0766 \\
$2$ 			&& 1.1171 && 0.2752 &&  0.9046 &&   1.5854 &&   0.0717 \\
$J\times(G-1)$  && 1.1098 && 0.2529 &&	0.9074 &&	1.6784 &&	0.1160 \\
$n/10$ 			&& 1.1070 && 0.2530 &&	0.9074 &&	1.6885 &&	0.1220 \\
$n/5$ 			&& 1.0670 && 0.2280 &&	0.9103 &&	1.7330 &&	0.1708 \\
$n/2$ 			&& 1.0185 && 0.2452 &&	0.9093 &&	1.8354 &&	0.2476 \\
$n/1.25$ 		&& 1.1507 && 0.3895 &&	0.9020 &&	1.9246 &&	0.3003 \\
$n/1.11$ 		&& 1.5037 && 0.5548 &&	0.8759 &&	1.9461 &&	0.3364 \\

\cmidrule{3-11}

&&\multicolumn{9}{c}{Mixing proportions $(0.1,0.2,0.3,0.4)$}\\
     &&	       &&        &&	   &&        &&        \\
$1$ 			&& 2.3342 && 	0.3511 && 	0.9210 && 	1.3371 && 	0.0265 \\
$2$ 			&& 2.2934 && 	0.3441 && 	0.9213 && 	1.3850 && 	0.0256 \\
$J\times(G-1)$  && 2.2529 &&	0.3521 &&	0.9222 &&	1.4513 &&	0.0347 \\
$n/10$ 			&& 2.2170 &&	0.3481 &&	0.9229 &&	1.4724 &&	0.0362 \\
$n/5$ 			&& 2.0861 &&	0.3066 &&	0.9290 &&	1.5641 &&	0.0553 \\
$n/2$ 			&& 1.9809 &&	0.2375 &&	0.9343 &&	1.7435 &&	0.1486 \\
$n/1.25$ 		&& 2.1211 &&	0.3672 &&	0.9235 &&	1.8396 &&	0.2866 \\
$n/1.11$ 		&& 2.3623 &&	0.4321 &&	0.9166 &&	1.8031 &&	0.3218 \\

\hline \hline 
\end{tabular}}
\caption{ConK. Results for different values of $k.$ 250 samples, $n=100,$ 10 random starts, 3 regressors and intercept. Values averaged across samples. \label{tab:100obskdel}}
\end{table}

\FloatBarrier

Tables \ref{tab:100obskdel} and \ref{tab:200obskdel} show results of the ConK algorithm for $n=100$ and $n=200$ respectively, for 8 different values of $k$. Whereby we observe some variation across conditions with $n=100$, results are qualitatively the same for $n=200$, except for the inadmissible $k=n/1.11$. For sake of conciseness, in subsequent analysis we focus on the perhaps most representative scenarios of $k=1$ and $k=n/5$, respectively \emph{small} and \emph{large} $k$.

In Tables \ref{tab:100obs} and \ref{tab:200obs} we display results for all approaches in terms of average MSE of model parameters, adj-Rand index, CPU time and selected $c$ for, respectively, $n=100$ and $n=200.$ Similarly to what Di Mari et al. (2017) found, with two components and $n=100,$ the difference between the unconstrained and the constrained approaches is small in terms of accuracy of regression parameter estimates. With $G=3$ and $G=4$ the gain of using a constrained approach is neater, especially with uneven mixing proportions. We observe that, together with keeping up very well with the cross-validated constrained approach, the proposed accelerated method performs equally well given the two alternative $k$'s, and does from twice up to nearly ten times faster than ConC. Interestingly, ConK$_{k=n/5}$'s solutions are closer to ConC in terms of selected scale balance than ConK$_{k=1}$'s. In terms of clusters recovery, all of the three constrained setups give better results compared to the unconstrained algorithms, and are relatively close to one another. None of the three constrained approaches does systematically better than the other two, although it seems that the cross-validation based method tends to do better both in terms of MSE of regression parameters and adj-Rand index.

Considering a sample size of $n=200$ (Table \ref{tab:200obs}) boosts the performance of all methods, especially of the constrained methods while HetN and HomN show a good improvement in the $(0.25,0.25,0.25,0.25)'$ condition only. The conditions where $n=200,$ $G=3$ and $G=4$ is where the difference between the constrained and the unconstrained methods is largest - with an average MSE for the estimated regressors and the component variances at least 50 \% smaller. In all conditions ConC and ConK deliver the best clusters recovery.

\begin{table}
\centering
\resizebox{0.48\vsize}{!}{\begin{tabular}{lcccccccccc}
\hline \hline
$k$ && Avg MSE $\hat{\pmb{\beta}}$  && Avg MSE $\hat{\pmb{\sigma}}^{2}$ && $\text{Adj-Rand}$ && time && $c$ \\
 \hline

&&\multicolumn{9}{c}{Mixing proportions $(0.5,0.5)$}\\
     &&	       &&        &&	   &&        &&        \\
$1$ 			&& 0.0057 &&	0.0032 && 	0.9784 && 	0.2449 && 	0.3427 \\
$2$ 			&& 0.0057 && 	0.0032 && 	0.9783 && 	0.2535 && 	0.3728 \\   	
$J\times(G-1)$  && 0.0057 &&	0.0032 &&	0.9783 &&	0.2551 && 	0.3886 \\
$n/10$ 			&& 0.0057 &&	0.0033 &&	0.9778 &&	0.3590 &&	0.6057 \\
$n/5$ 			&& 0.0058 &&	0.0051 &&	0.9754 &&	0.3820 &&	0.7022 \\
$n/2$ 			&& 0.0059 &&	0.0072 &&	0.9746 &&	0.3824 &&	0.6826 \\
$n/1.25$ 		&& 0.0058 &&	0.0053 &&	0.9764 &&	0.3683 &&	0.6066 \\
$n/1.11$ 		&& 0.0058 &&	0.0050 &&	0.9767 &&	0.3628 &&	0.5962 \\

\cmidrule{3-11}

&&\multicolumn{9}{c}{Mixing proportions $(0.2,0.8)$}\\
     &&	       &&        &&	   &&        &&        \\
$1$ 			&& 0.0110 && 	0.0062 && 	0.9842 && 	0.2435 && 	0.2587 \\
$2$ 			&& 0.0110 && 	0.0062 && 	0.9842 && 	0.2471 && 	0.2610 \\
$J\times(G-1)$ 	&& 0.0110 &&	0.0062 &&	0.9842 &&	0.2498 &&	0.2585 \\
$n/10$ 			&& 0.0109 &&	0.0061 &&	0.9841 &&	0.2521 &&	0.2650 \\
$n/5$ 			&& 0.0109 &&	0.0061 &&	0.9842 &&	0.2567 &&	0.2676 \\
$n/2$ 			&& 0.0111 &&	0.0081 &&	0.9839 &&	0.2645 &&	0.2866 \\
$n/1.25$ 		&& 0.0111 &&	0.0071 &&	0.9820 &&	0.3383 &&	0.4848 \\
$n/1.11$ 		&& 0.0112 &&	0.0124 &&	0.9809 &&	0.3515 &&	0.5578 \\

\cmidrule{3-11}

&&\multicolumn{9}{c}{Mixing proportions $(0.2,0.4,0.4)$}\\
     &&	       &&        &&	   &&        &&        \\
$1$ 			&& 0.0103 && 	0.0067 && 	0.9856 && 	0.7080 && 	0.1308 \\
$2$ 			&& 0.0103 && 	0.0067 && 	0.9856 && 	0.7127 && 	0.1302 \\
$J\times(G-1)$ 	&& 0.0103 &&	0.0067 &&	0.9857 &&	0.7444 &&	0.1714  \\
$n/10$ 			&& 0.0103 &&	0.0070 &&	0.9860 &&	0.8391 &&	0.2577  \\
$n/5$ 			&& 0.0104 &&	0.0086 &&	0.9855 &&	0.9152 &&	0.3064 \\
$n/2$ 			&& 0.0104 &&	0.0142 &&	0.9846 &&	0.9816 &&	0.3340 \\
$n/1.25$ 		&& 0.0111 &&	0.0363 &&	0.9832 &&	0.9338 &&	0.3765 \\
$n/1.11$ 		&& 0.0329 &&	0.0649 &&	0.9809 &&	0.9117 &&	0.3750 \\

\cmidrule{3-11}

&&\multicolumn{9}{c}{Mixing proportions $(0.2,0.3,0.5)$}\\
     &&	       &&        &&	   &&        &&        \\
$1$ 			&& 0.0152 && 0.0107 && 0.9865 && 0.8306 && 0.0991 \\
$2$ 			&& 0.0152 && 0.0106 && 0.9865 && 0.8429 && 0.0908 \\
$J\times(G-1)$ 	&& 0.0152 && 0.0106 && 0.9864 && 0.8588 && 0.1114 \\
$n/10$ 			&& 0.0152 && 0.0106 && 0.9870 && 0.9706 && 0.1441 \\
$n/5$ 			&& 0.0152 && 0.0112 && 0.9868 && 0.9826 && 0.1672 \\
$n/2$ 			&& 0.0104 && 0.0190 && 0.9860 && 1.0970 && 0.2805  \\
$n/1.25$ 		&& 0.0110 && 0.0752 && 0.9834 && 1.1029 && 0.3326 \\
$n/1.11$ 		&& 0.0318 && 0.1397 && 0.9787 && 1.1643 && 0.3610 \\

\cmidrule{3-11}

&&\multicolumn{9}{c}{Mixing proportions $(0.25,0.25,0.25,0.25)$}\\
     &&	       &&        &&	   &&        &&        \\
$1$				&& 0.0834 && 	0.0289 && 	0.9804 && 	2.1388 && 	0.1089 \\
$2$ 			&& 0.0901 && 	0.0361 && 	0.9782 && 	2.1607 && 	0.1116 \\
$J\times(G-1)$ 	&& 0.0901 &&	0.0358 &&	0.9781 &&	2.2245 &&	0.1790 \\
$n/10$			&& 0.0969 &&	0.0365 &&	0.9780 &&	2.3181 &&	0.2546 \\
$n/5$ 			&& 0.0964 &&	0.0390 &&	0.9775 &&	2.3737 &&	0.3376  \\
$n/2$ 			&& 0.1060 &&	0.0487 &&	0.9760 &&	2.4668 &&	0.4139 \\
$n/1.25$		&& 0.1752 &&	0.0877 &&	0.9692 &&	2.5395 &&	0.3217 \\
$n/1.11$ 		&& 0.5128 &&	0.1659 &&	0.9521 &&	2.4057 &&	0.3110 \\

\cmidrule{3-11}

&&\multicolumn{9}{c}{Mixing proportions $(0.1,0.2,0.3,0.4)$}\\
     &&	       &&        &&	   &&        &&        \\
$1$ 			&& 0.2253 && 	0.0648 && 	0.9845 && 	2.5761 && 	0.0357 \\
$2$ 			&& 0.2351 && 	0.0643 && 	0.9844 && 	2.6548 && 	0.0302 \\
$J\times(G-1)$ 	&& 0.1941 &&	0.0509 &&	0.9870 &&	2.7378 &&	0.0387 \\
$n/10$ 			&& 0.1861 &&	0.0505 &&	0.9871 &&	2.9975 &&	0.0548 \\
$n/5$ 			&& 0.1637 &&	0.0433 &&	0.9880 &&	3.1908 &&	0.0769 \\
$n/2$ 			&& 0.1753 &&	0.0595 &&	0.9863 &&	3.6008 &&	0.1459 \\
$n/1.25$ 		&& 0.1712 &&	0.1572 &&	0.9852 &&	3.6715 &&	0.2814 \\
$n/1.11$ 		&& 0.2562 &&	0.2362 &&	0.9811 &&	4.1774 &&	0.3322 \\

\hline \hline 
\end{tabular}}
\caption{ConK. Results for different values of $k.$ 250 samples, $n=200,$ 10 random starts, 3 regressors and intercept. Values averaged across samples. \label{tab:200obskdel}}
\end{table}

\FloatBarrier

\begin{table}
\centering
\begin{tabular}{lcccccccccc}
\hline \hline
Algorithm && Avg MSE $\hat{\pmb{\beta}}$  && Avg MSE $\hat{\pmb{\sigma}}^{2}$ && $\text{Adj-Rand}$ && time && $c$ \\
 \hline

&&\multicolumn{9}{c}{Mixing proportions $(0.5,0.5)$}\\
     &&	       &&        &&	   &&        &&        \\     
	
HomN && 0.0159 && 0.0367 && 0.9399 && 0.0485 &&   -    \\
HetN && 0.0135 && 0.0076 && 0.9616 && 0.0492 &&   -    \\
ConC && 0.0135 && 0.0082 && 0.9626 && 0.7694 && 0.4824 \\
Con$\text{K}_{k=1}$ && 0.0135 && 0.0075 && 0.9619 && 0.1184 && 0.3024 \\
Con$\text{K}_{k=n/5}$ && 0.0136 &&	0.0086 &&	0.9613 &&	0.1765 &&	0.6284 \\

\cmidrule{3-11}

&&\multicolumn{9}{c}{Mixing proportions $(0.2,0.8)$}\\
     &&	       &&        &&	   &&        &&        \\

HomN && 0.0402 && 0.0510 && 0.9648 && 0.0567 &&  -     \\
HetN && 0.0522 && 0.0439 && 0.9672 && 0.0525 &&  -     \\
ConC && 0.0416 && 0.0238 && 0.9718 && 0.8026 && 0.5102 \\
Con$\text{K}_{k=1}$ && 0.0365 && 0.0152 && 0.9726 && 0.1533 && 0.2353 \\
Con$\text{K}_{k=n/5}$ && 0.0364 &&	0.0152 &&	0.9729 &&	0.1534 &&	0.2456 \\

\cmidrule{3-11}

&&\multicolumn{9}{c}{Mixing proportions $(0.2,0.4,0.4)$}\\
     &&	       &&        &&	   &&        &&        \\

HomN && 0.3192 && 0.1107 && 0.9403 && 0.1763 &&  -     \\
HetN && 0.3201 && 0.1665 && 0.9318 && 0.1804 &&  -     \\
ConC && 0.2117 && 0.0916 && 0.9481 && 1.4315 && 0.2875 \\
Con$\text{K}_{k=1}$ && 0.2009 && 0.0691 && 0.9517 && 0.5240 && 0.1201 \\
Con$\text{K}_{k=n/5}$ && 0.2084 && 0.0780 && 0.9465 && 0.7067 && 0.2430 \\

\cmidrule{3-11}

&&\multicolumn{9}{c}{Mixing proportions $(0.2,0.3,0.5)$}\\
     &&	       &&        &&	   &&        &&        \\

HomN && 0.2482 && 0.0899 && 0.9546 && 0.1868 &&   -    \\
HetN && 0.3226 && 0.1226 && 0.9579 && 0.1446 &&   -    \\
ConC && 0.1051 && 0.0462 && 0.9688 && 1.2933 && 0.2436 \\
Con$\text{K}_{k=1}$ && 0.1256 && 0.0450 && 0.9690 && 0.4710 && 0.0929 \\
Con$\text{K}_{k=n/5}$ && 0.0934 && 0.0397 &&	0.9696 && 0.5413 &&	0.1820 \\

\cmidrule{3-11}

&&\multicolumn{9}{c}{Mixing proportions $(0.25,0.25,0.25,0.25)$}\\
     &&	       &&        &&	   &&        &&        \\

HomN && 1.1263 && 0.2242 && 0.8800 && 0.3373 &&  -     \\
HetN && 2.5668 && 0.6200 && 0.8239 && 0.3349 &&  -     \\
ConC && 1.0796 && 0.2735 && 0.9069 && 2.6223 && 0.1704 \\
Con$\text{K}_{k=1}$ && 1.0741 && 0.2464 && 0.9083 && 1.5461 && 0.0766 \\
Con$\text{K}_{k=n/5}$ && 1.0670 && 0.2280 &&	0.9103 &&	1.7330 &&	0.1708 \\

\cmidrule{3-11}

&&\multicolumn{9}{c}{Mixing proportions $(0.1,0.2,0.3,0.4)$}\\
     &&	       &&        &&	   &&        &&        \\

HomN && 3.8816 && 0.2174 && 0.9115 && 0.3384 &&   -    \\
HetN && 3.8681 && 0.6046 && 0.8802 && 0.3120 &&   -    \\
ConC && 2.5493 && 0.3579 && 0.9193 && 2.5098 && 0.0591 \\
Con$\text{K}_{k=1}$ && 2.3342 && 0.3511 && 0.9210 && 1.3371 && 0.0265 \\
Con$\text{K}_{k=n/5}$ && 2.0861 &&	0.3066 &&	0.9290 &&	1.5641 &&	0.0553 \\
\hline \hline 
\end{tabular}
\caption{250 samples, $n=100,$ 10 random starts, 3 regressors and intercept. Values averaged across samples. \label{tab:100obs}}
\end{table}

\begin{table}
\centering
\begin{tabular}{lcccccccccc}
\hline \hline
Algorithm && Avg MSE $\hat{\pmb{\beta}}$  && Avg MSE $\hat{\pmb{\sigma}}^{2}$ && $\text{Adj-Rand}$ && time && $c$ \\
 \hline

&&\multicolumn{9}{c}{Mixing proportions $(0.5,0.5)$}\\
     &&	       &&        &&	   &&        &&        \\     
	
HomN && 0.0070 && 0.0284 && 0.9574 && 0.0954 &&    -   \\
HetN && 0.0057 && 0.0032 && 0.9784 && 0.0977 &&    -   \\
ConC && 0.0057 && 0.0033 && 0.9777 && 2.2902 && 0.4884 \\
Con$\text{K}_{k=1}$ && 0.0057 && 0.0032 && 0.9784 && 0.2449 && 0.3427 \\
Con$\text{K}_{k=n/5}$ && 0.0058 &&	0.0051 &&	0.9754 &&	0.3820 &&	0.7022 \\

\cmidrule{3-11}
%\hline

&&\multicolumn{9}{c}{Mixing proportions $(0.2,0.8)$}\\
     &&	       &&        &&	   &&        &&        \\

HomN && 0.0121 && 0.0384 && 0.9762 && 0.1116 &&   -    \\
HetN && 0.0137 && 0.0083 && 0.9839 && 0.0954 &&   -    \\
ConC && 0.0109 && 0.0059 && 0.9841 && 2.2682 && 0.4633 \\
Con$\text{K}_{k=1}$ && 0.0110 && 0.0062 && 0.9842 && 0.2435 && 0.2587 \\
Con$\text{K}_{k=n/5}$ && 0.0109 &&	0.0061 &&	0.9842 &&	0.2567 &&	0.2676 \\

\cmidrule{3-11}
%\hline

&&\multicolumn{9}{c}{Mixing proportions $(0.2,0.4,0.4)$}\\
     &&	       &&        &&	   &&        &&        \\

HomN && 0.0553 && 0.0724 && 0.9716 && 0.3251 &&   -    \\
HetN && 0.0399 && 0.0208 && 0.9835 && 0.2951 &&   -    \\
ConC && 0.0103 && 0.0075 && 0.9856 && 3.4715 && 0.2768 \\
Con$\text{K}_{k=1}$ && 0.0103 && 0.0067 && 0.9856 && 0.7080 && 0.1308 \\
Con$\text{K}_{k=n/5}$ && 0.0104 &&	0.0086 &&	0.9855 &&	0.9152 &&	0.3064 \\

\cmidrule{3-11}
%\hline

&&\multicolumn{9}{c}{Mixing proportions $(0.2,0.3,0.5)$}\\
     &&	       &&        &&	   &&        &&        \\

HomN && 0.3245 && 0.1476 && 0.9637 && 0.4227 &&        \\
HetN && 0.0630 && 0.0234 && 0.9838 && 0.2868 &&        \\
ConC && 0.0152 && 0.0116 && 0.9862 && 3.6266 && 0.2067 \\
Con$\text{K}_{k=1}$ && 0.0152 && 0.0107 && 0.9865 && 0.8306 && 0.0991 \\
Con$\text{K}_{k=n/5}$ && 0.0152 && 0.0112 && 0.9868 && 0.9826 && 0.1672 \\

\cmidrule{3-11}

&&\multicolumn{9}{c}{Mixing proportions $(0.25,0.25,0.25,0.25)$}\\
     &&	       &&        &&	   &&        &&        \\

HomN && 0.0355 && 0.0562 && 0.9754 && 0.6611 &&  -     \\
HetN && 0.5711 && 0.1667 && 0.9482 && 0.7157 &&  -     \\
ConC && 0.1213 && 0.0376 && 0.9777 && 5.6341 && 0.2019 \\
Con$\text{K}_{k=1}$ && 0.0834 && 0.0289 && 0.9804 && 2.1388 && 0.1089 \\
Con$\text{K}_{k=n/5}$ && 0.0964 &&	0.0390 &&	0.9775 &&	2.3737 &&	0.3376  \\

\cmidrule{3-11}

&&\multicolumn{9}{c}{Mixing proportions $(0.1,0.2,0.3,0.4)$}\\
     &&	       &&        &&	   &&        &&        \\

HomN && 3.0294 && 0.2477 && 0.9484 && 0.7421 &&   -    \\
HetN && 0.9792 && 0.2242 && 0.9686 && 0.6644 &&   -    \\
ConC && 0.1865 && 0.0481 && 0.9871 && 6.3252 && 0.0655 \\
Con$\text{K}_{k=1}$ && 0.2253 && 0.0648 && 0.9845 && 2.5761 && 0.0357 \\
Con$\text{K}_{k=n/5}$ && 0.1637 &&	0.0433 &&	0.9880 &&	3.1908 &&	0.0769 \\

\hline \hline 
\end{tabular}
\caption{250 samples, $n=200,$ 10 random starts, 3 regressors and intercept. Values averaged across samples. \label{tab:200obs}}
\end{table}

\FloatBarrier

In Table \ref{tab:modseltab} we display the percentage of correct guesses for $G$ delivered by each method for each of the 12 simulation conditions. The procedure minimizing the BIC computed using the solutions of HetN almost completely fails to recover the correct number of clusters. By contrast, we observe that in all conditions the modified BIC computed using the constrained approaches yields the highest number of correct guesses. Very similar performance is achieved with standard BIC computed at the ConC solution. Using standard BIC tarnishes the performance of ConK, which however outperforms HomN in setups with larger sample sizes ($n=200$), as well as with smaller sample size ($n=100$) but larger $G^*$ and uneven component sizes.  

\begin{figure}
\centering
	\includegraphics[scale=0.45]{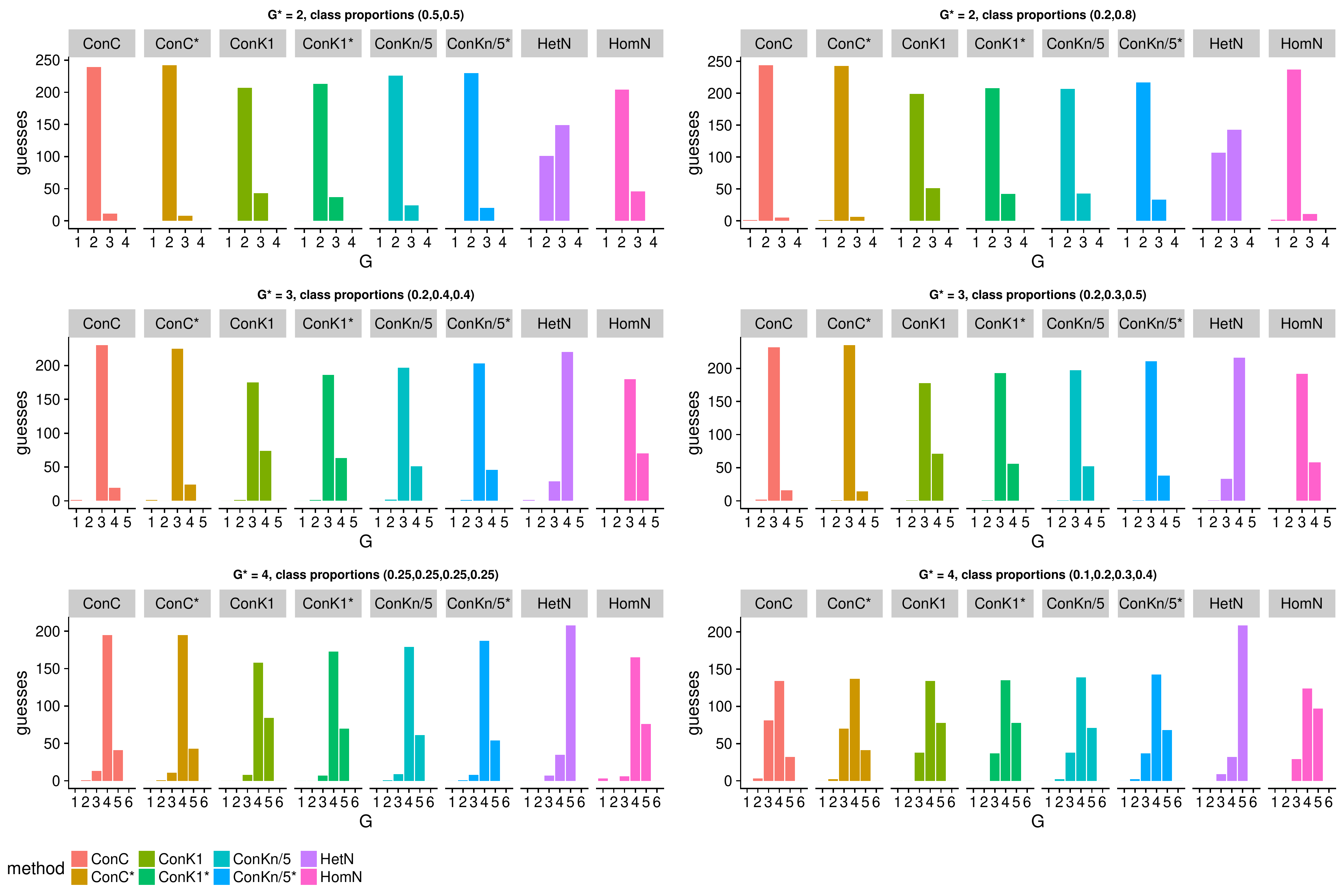}
	\caption{Number of guesses for $G$ per method, $n=100.$
	 The cross-validation procedure is run with $M = n/5$ and test set of size = $n/10.$}
	\label{fig:mods10p55}
\end{figure}

\begin{figure}
\centering
	\includegraphics[scale=0.45]{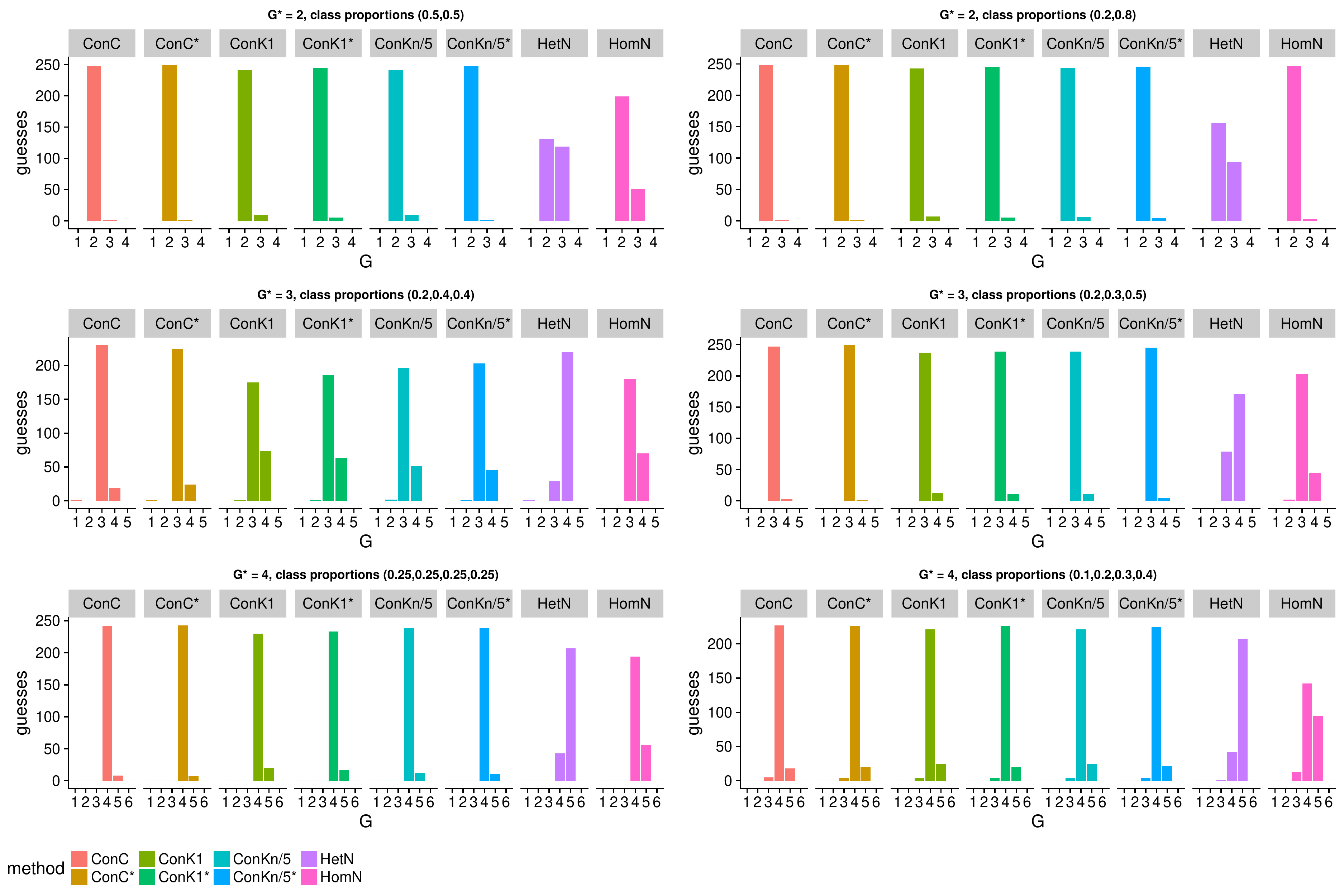}
	\caption{Number of guesses for $G$ per method, $n=200.$ 
	 The cross-validation procedure is run with $M = n/5$ and test set of size = $n/10.$}
	\label{fig:mods20p55}
\end{figure}

Further insight can be acquired by looking at the absolute frequencies of guesses for the number of components $G$ of each method (Figure \ref{fig:mods10p55} and \ref{fig:mods20p55}). For the constrained approaches ConC and ConK, we compute the BIC based on the formulas of Equations \eqref{eq:constrbic} and \eqref{eq:bicCer}. The advantage of ConK and ConC over HomN - which does slightly better than HetN - shows up in all conditions, and is more evident for $n=200$. We observe that the correction for the different model complexities entailed by $c$ shows an improvement in selecting the number of components for $n = 100$, which vanishes for ConC when $n=200$, and is neater for ConK\footnote{We checked that this is also the case for $n>200$. Related figures are available from the corresponding author upon request}. One possible explanation for this might be that the cross-validation strategy for selecting the scale balance is less affected by overspecification of the number of components - although when the number of components is well specified, ConK delivers almost no loss relative to ConC in parameter estimation.  
\begin{table}
\centering
\resizebox{1\hsize}{!}{\begin{tabular}{lcccccccccccc}
\hline \hline
$n=100$     &&	\multicolumn{3}{c}{$G=2$}   &&	\multicolumn{3}{c}{$G=3$}   && \multicolumn{3}{c}{$G=4$} \\
\cmidrule{3-5} \cmidrule{7-9} \cmidrule{11-13} 
  && (0.5,0.5)'  && (0.2,0.8)' && (0.2,0.4,0.4)' && (0.2,0.3,0.5)' && (0.25,0.25,0.25,0.25)' && (0.1,0.2,0.3,0.4)' \\
     
     \cmidrule{3-13}
HomN && 0.816 && 0.948 && 0.720 && 0.768 && 0.660 && 0.496 \\
HetN && 0.404 && 0.428 && 0.116 && 0.132 && 0.140 && 0.128 \\
ConC && 0.956 && 0.976 && 0.920 && 0.928 && 0.780 && 0.536 \\
ConC$^*$  && 0.968 && 0.972 && 0.900 && 0.940 && 0.780 && 0.548 \\
ConK$_{k=1}$  && 0.828 && 0.796 && 0.700 && 0.712 && 0.632 && 0.536 \\
ConK$_{k=1}^*$ && 0.852 && 0.832 && 0.744 && 0.772 && 0.692 && 0.540 \\
ConK$_{k=n/5}$ && 0.832 && 0.800 && 0.704 && 0.712 && 0.652 && 0.532 \\
ConK$_{k=n/5}^*$ && 0.860 && 0.832 && 0.748 && 0.776 && 0.700 && 0.536 \\
\hline  
$n=200$     &&	\multicolumn{3}{c}{$G=2$}   &&	\multicolumn{3}{c}{$G=3$}   && \multicolumn{3}{c}{$G=4$} \\
\cmidrule{3-5} \cmidrule{7-9} \cmidrule{11-13} 
  && (0.5,0.5)'  && (0.2,0.8)' && (0.2,0.4,0.4)' && (0.2,0.3,0.5)' && (0.25,0.25,0.25,0.25)' && (0.1,0.2,0.3,0.4)' \\
     
     \cmidrule{3-13}
HomN && 0.796 && 0.988 && 0.804 && 0.812 && 0.776 && 0.568 \\
HetN && 0.524 && 0.624 && 0.300 && 0.316 && 0.172 && 0.168 \\
ConC && 0.992 && 0.992 && 0.980 && 0.988 && 0.968 && 0.908 \\
ConC$^*$  && 0.996 && 0.992 && 0.988 && 0.996 && 0.972 && 0.904 \\
ConK$_{k=1}$ && 0.964 && 0.972 && 0.944 && 0.948 && 0.920 && 0.884 \\
ConK$_{k=1}^*$ && 0.980 && 0.980 && 0.968 && 0.956 && 0.932 && 0.904 \\
 ConK$_{k=n/5}$ && 0.964 && 0.972 && 0.944 && 0.944 && 0.916 && 0.876 \\
ConK$_{k=n/5}^*$ && 0.980 && 0.984 && 0.968 && 0.956 && 0.932 && 0.896 \\
\hline  
     
     \end{tabular}}
\caption{Proportion of correct guesses for $G$. 250 samples, 10 random starts, 3 regressors and intercept. Values averaged across samples. For each setup, lowest BIC solution selected between $1,\dots,G^{*}+2$ components. Methods with index $^*$ use the BIC formula of Equation \eqref{eq:bicCer}. \label{tab:modseltab}}     
     \end{table} 

\FloatBarrier     
     
\FloatBarrier

\begin{figure}
\centering
	\includegraphics[scale=0.45]{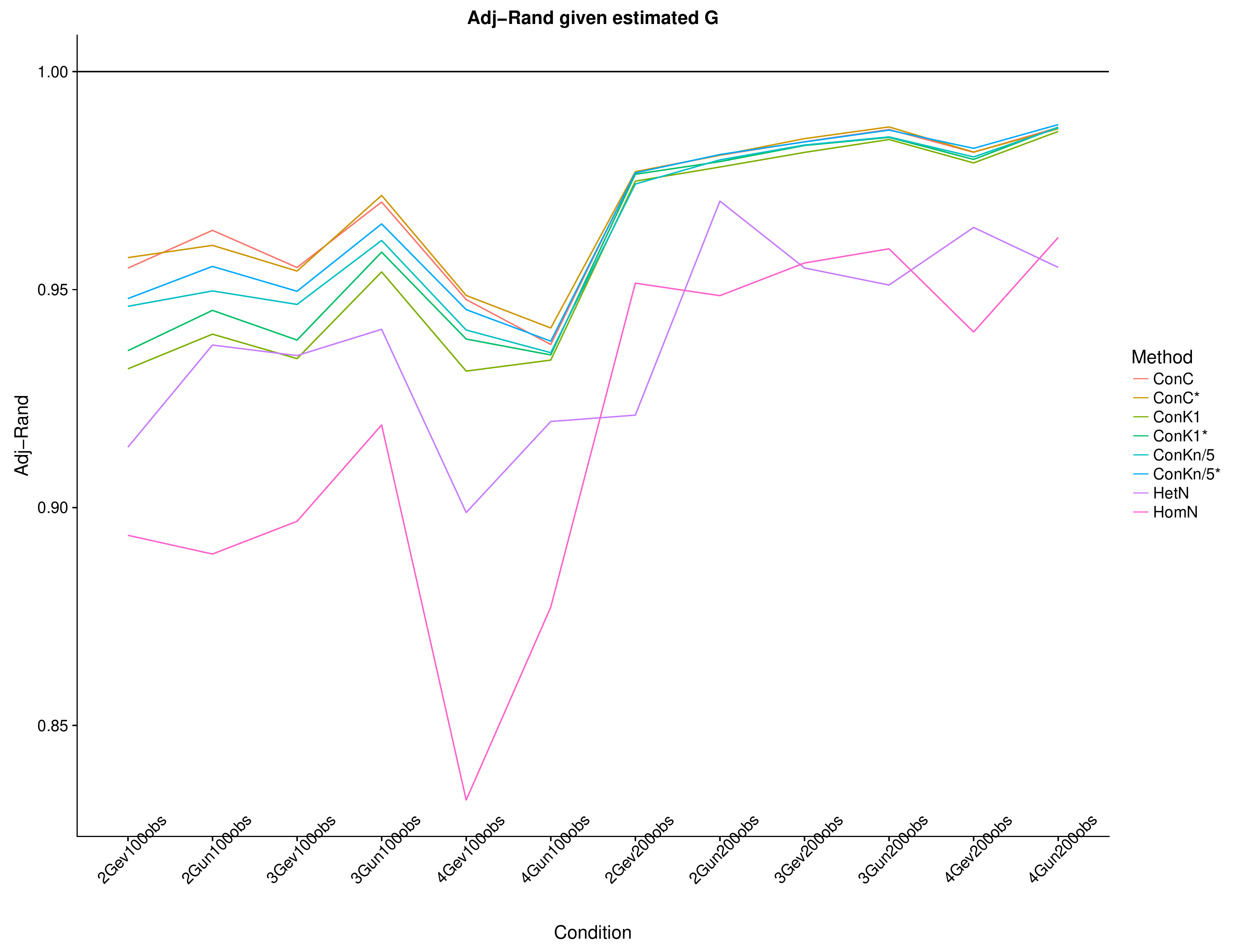}
	\caption{Average Adjusted Rand Index (Adj-Rand) for each method, in all 12 simulation conditions, with the optimal $G$ as selected by each method.}
	\label{fig:arandg}
\end{figure}

Finally, in Figure \ref{fig:arandg} we plot the average Adjusted Rand index, computed using the solution with number of components as chosen by each method. For instance, if HetN has chosen 4 components when $G^*=2$, we compute the Adj-Rand comparing the 4-component solution with the true 2-component one. Overall, by simultaneously looking at model selection and cluster recovery, all the constrained methods, with and without the correction for the number of degrees of freedom, yield very similar performances, doing always better than the unconstrained rivals. 

\FloatBarrier

\section{Two real-data applications}\label{sec:data} 
In this Section we illustrate the use of the constrained approaches, ConC and ConK, and compare them with HetN and HomN. 

For neither of the data sets the number of subgroups in the underlying population is known. We fitted a clusterwise linear regression, using the 3 methods under comparison, on the \emph{CEO} data set (\url{http://lib.stat.cmu.edu/DASL/DataArchive.html}), with 2, 3, 4, and 5 components, and analyzed the 2-class solution - which minimized the BIC formulas of Equation \eqref{eq:constrbic} and \eqref{eq:bicCer} under ConC and ConK (for $k=1$ and $k=n/5$), and the plain BIC under HomN - in terms of estimated model parameters and clustering. We carried out a similar exercise on the \emph{AutoMpg} data set (available at \url{https://archive.ics.uci.edu/ml/machine-learning-databases/auto-mpg/auto-mpg.data}), where instead only the constrained methods agree on the 2-component solution - seemingly the most suited to the data in terms of clusters interpretation.

\subsection{\emph{CEO} data}

This data set contains information about salary (dependent variable) and age (independent variable) of 59 CEOs from small U.S. companies. The underlying clusters structure is unknown. Among those who already analyzed this data set, Carbonneau, Caporossi, and Hansen (2011) fitted a 2-component clusterwise linear regression, whereas Bagirov, Ugon, and Mirzayeva (2013) compared the 2-component and the 4-component setups.

In Table \ref{tab:BICceo} we show BIC values computed using respectively HomN and HetN, and BIC's (Equations \eqref{eq:constrbic} and \eqref{eq:bicCer}) computed using ConC and ConK (with $k=1$ and $k=n/5$). The constrained approaches and HomN all agree on the two component solution. Using BIC with HetN would lead to select the seemingly spurious 4-component solution showed in Di Mari et al. (2017).

\begin{table}[h!]
\centering
\begin{tabular}{lccccc}
\hline \hline
							&& $G=2$  & $G=3$  & $G=4$  & $G=5$  \\
							\cmidrule{3-6}
BIC$_{\text{HomN}}$ && {\bf 706.30} & 712.14 & 719.41 & 725.66 \\
BIC$_{\text{HetN}}$ && 704.42 & 707.70 & {\bf 593.94} & 601.88 \\
BIC$_{\text{ConC}}$ && {\bf 706.86} & 721.13 & 740.92 & 741.53 \\
BIC$_{\text{ConC$^*$}}$ && {\bf 702.12} & 716.45 & 724.61 & 725.10 \\
BIC$_{\text{ConK$_{k=1}$}}$  && {\bf 704.42} & 707.72 & 712.01 & 718.95 \\
BIC$_{\text{ConK$_{k=1}^*$}}$ && {\bf 703.23} & 707.72 & 712.01 & 718.95 \\
BIC$_{\text{ConK$_{k=n/5}$}}$ && {\bf 707.25} & 713.76 & 727.35 & 727.88 \\
BIC$_{\text{ConK$_{k=n/5}^*$}}$ && {\bf 702.12} & 713.71 & 721.95 & 727.76 \\

\hline \hline
\end{tabular}
\caption{\emph{CEO} data. BIC values for $G=2$, $G=3$, $G=4$, and $G=5$. Best solutions out of 100 random starts. Minimum BIC values in bold for each method. \label{tab:BICceo}}
\end{table}

\clearpage
\begin{center}
\begin{figure}[h!]
	\includegraphics[scale=1]{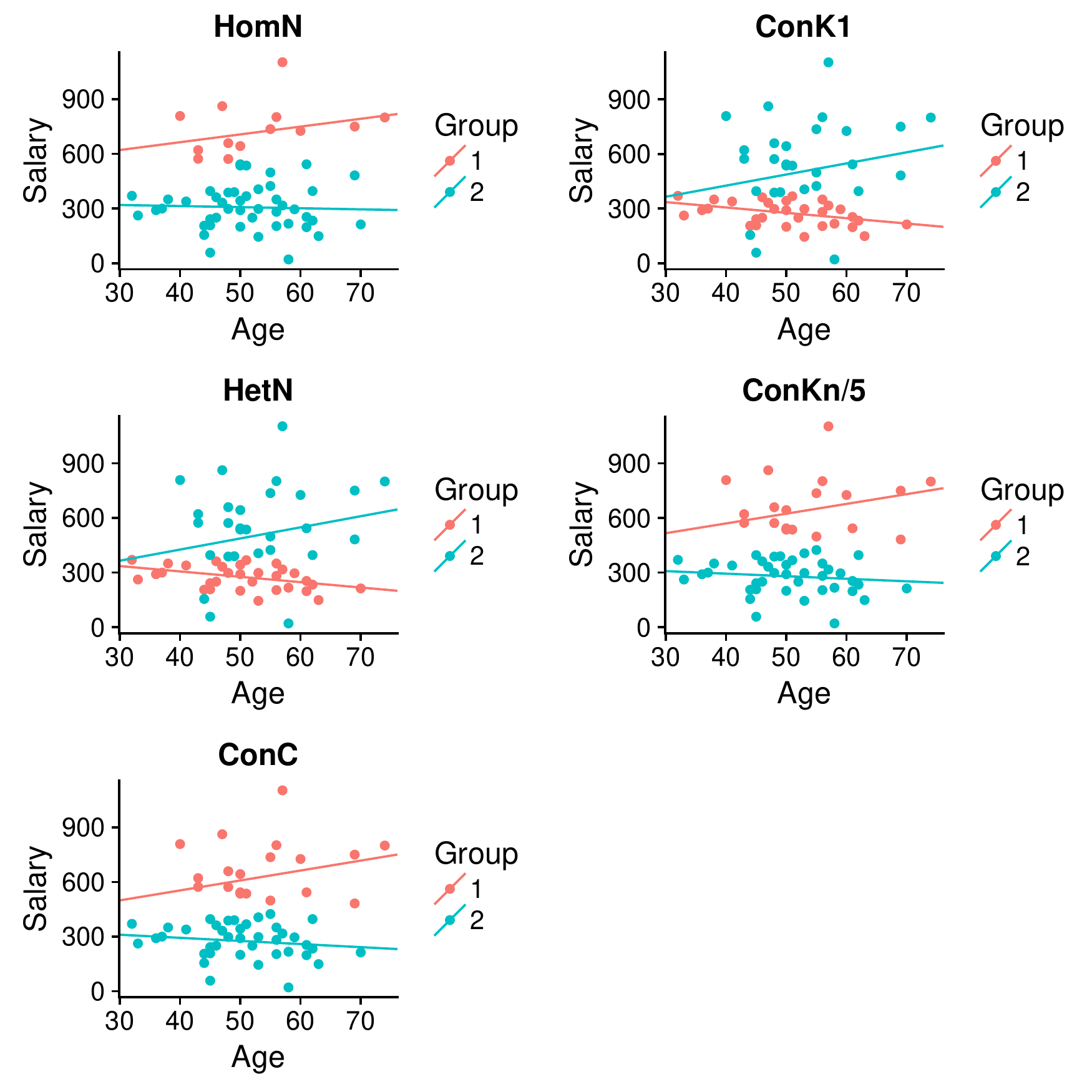}
	\caption{\emph{CEO} data. Clusterwise regressions of salary on age of CEO's. Best solutions out of 100 random starts, $G= 2$.
	 The cross-validation procedure is run with $M = n/5$ and test set of size = $n/10.$}
	\label{fig:ceo2}
\end{figure}
\end{center}
\FloatBarrier

The 2-class clusterwise linear regressions and the crisp assignments are plotted in Figure \ref{fig:ceo2}. 

We observe that, in line with the simulation study, ConC and ConK with $k=n/5$ yield the same clustering, and such clustering is exactly in between HomN and HetN, as well as very similar regression lines. This confirms what was found in Di Mari et al. (2017). By contrast, ConK with $k=1$ produces a final solution which is closer to HetN.

\subsection{\emph{AutoMpg} data}
This data set contains a sample of 398 vehicles, where information on city-cycle fuel consumption in miles per gallon is gathered for each vehicle, alongside with the following set of covariates (of mixed type): number of cylinders, model year, and origin, which are discrete valued; displacement, horsepower, weight, and acceleration, which are instead continuous valued. Records for horsepower were missing for six sample units. Given that the car model is available, with all relevant information, we were able to retrieve the missing values and included them in the data set.

We estimated a clusterwise linear regression model of miles per gallon on the above set of covariates. Plain BIC and modified BIC - for ConC and ConK only - values are reported in Table \ref{tab:BICautompg}. Constrained approaches largely agree on the two-component solution (5 out of 6), whereby HomN and HetN favor respectively the 3 and the 5 component solution. Interestingly, ConK with $k=1$ seems to need a correction for model complexity in the BIC to behave coherently with ConC and ConK with $k=1.$

\FloatBarrier

\begin{table}[h!]
\centering
\begin{tabular}{lccccc}
\hline \hline
							&& $G=2$  & $G=3$  & $G=4$  & $G=5$  \\
							\cmidrule{3-6}
BIC$_\text{HomN}$ && 1365.89 & {\bf 1363.22} & 1381.12 & 1398.56 \\
BIC$_\text{HetN}$ && 1329.35 & 1340.55 & 1320.86 & {\bf 1319.93} \\
BIC$_\text{ConC}$ && {\bf 1329.74} & 1340.95 & 1364.76 & 1371.26 \\
BIC$_\text{ConC$^*$}$  && {\bf 1325.03} & 1336.11 & 1364.10 & 1370.17 \\
BIC$_\text{ConK$_{k=1}$}$  && 1329.35 & 1340.55 & {\bf 1328.49} & 1340.18 \\
BIC$_\text{ConK$_{k=1}^*$}$ && {\bf 1326.52} & 1338.08 & 1328.49 & 1340.13 \\
BIC$_\text{ConK$_{k=n/5}$}$ && {\bf 1337.27} & 1356.77 & 1361.80 & 1397.75 \\
BIC$_\text{ConK$_{k=n/5}^*$}$ && {\bf 1330.49} & 1344.93 & 1361.13 & 1389.29 \\

\hline \hline
\end{tabular}
\caption{\emph{Auto-Mpg} data. BIC values for $G=2$, $G=3$, $G=4$, and $G=5$. Best solutions out of 100 random starts. Minimum BIC values in bold for each method. \label{tab:BICautompg}}
\end{table}    
\FloatBarrier

By looking at Table \ref{tab:resautompg2} we observe that acceleration ($\mathbf{x}_1$), cylinders ($\mathbf{x}_2$), and displacement ($\mathbf{x}_3$) have all positive effect on miles per gallon in the first (smaller) component, and negative effect in the second (larger) component. Cars with more horsepower, not surprisingly, tend to drive less miles per gallon - although the effect is relatively milder for the second (larger) component - whereby a more recent model year ($\mathbf{x}_5$), all else equal, is positively associated with miles per gallon in both components - again, with a relatively milder effect on the second component.    
\begin{table}[h!]
\centering
\resizebox{\textwidth}{!}{\begin{tabular}{lccccccccccccccc}
\hline \hline
&&\multicolumn{2}{c}{HomN}&&\multicolumn{2}{c}{HetN}&&\multicolumn{2}{c}{ConC}&&\multicolumn{2}{c}{ConK$_{k=1}$} && \multicolumn{2}{c}{ConK$_{k=n/5}$}\\
\cmidrule{3-4} \cmidrule{6-7} \cmidrule{9-10} \cmidrule{12-13} \cmidrule{15-16}
$p_g$ &&   0.2215 &   0.7785 &&   0.4473 &   0.5527 &&   0.4353 &   0.5647 &&   0.4473 &   0.5527 &&   0.3857 &   0.6143 \\
intercept && -35.0716 &  -3.2278 && -23.3485 &   3.7071 && -23.5883 &   3.5861 && -23.3480 &   3.7072 && -24.7003 &   2.5239 \\
$\beta_{1g}$ &&   0.1819 &  -0.2530 &&   0.1354 &  -0.4212 &&   0.1383 &  -0.4177 &&   0.1354 &  -0.4212 &&   0.1601 &  -0.3867 \\
$\beta_{2g}$ &&   1.1272 &  -0.7172 &&   0.1853 &  -0.9055 &&   0.1767 &  -0.8938 &&   0.1853 &  -0.9055 &&   0.2145 &  -0.8550 \\
$\beta_{3g}$ &&   0.0170 &   0.0004 &&   0.0362 &  -0.0116 &&   0.0367 &  -0.0116 &&   0.0362 &  -0.0116 &&   0.0376 &  -0.0112 \\
$\beta_{4g}$ &&  -0.2113 &  -0.0077 &&  -0.1188 &  -0.0035 &&  -0.1211 &  -0.0031 &&  -0.1188 &  -0.0035 &&  -0.1286 &  -0.0015 \\
$\beta_{5g}$ &&   1.1328 &   0.5862 &&   0.9546 &   0.4699 &&   0.9592 &   0.4730 &&   0.9546 &   0.4699 &&   0.9773 &   0.4921 \\
$\beta_{6g}$ &&  -0.0070 &  -0.0042 &&  -0.0084 &  -0.0022 &&  -0.0084 &  -0.0022 &&  -0.0084 &  -0.0022 &&  -0.0085 &  -0.0026 \\
$\beta_{7g}$ &&   0.6887 &   1.7958 &&   0.7283 &   2.6872 &&   0.7221 &   2.6430 &&   0.7283 &   2.6872 &&   0.6703 &   2.4315 \\
$\sigma^2_g$ &&   2.3770 &   2.3770 &&   3.1592 &   1.4190 &&   3.1576 &   1.4906 &&   3.1592 &   1.4190 &&   3.1588 &   1.7886 \\
$c$ &&   - &   - &&   - &  - &&   0.1547 &   0.1547 &&   0.0558 &   0.0558 &&   0.3206 &   0.3206 \\

\hline \hline 

\end{tabular}}
\caption{\emph{Auto-Mpg} data. Covariates are acceleration ($\mathbf{x}_1$), cylinders ($\mathbf{x}_2$), displacement ($\mathbf{x}_3$), horsepower ($\mathbf{x}_4$), model year ($\mathbf{x}_5$), weight ($\mathbf{x}_6$), and origin ($\mathbf{x}_7$). Best solutions out of 100 random starts, $G= 2$. $K = n/5$, and test set size = $n/10.$ \label{tab:resautompg2}}
\end{table}

\section{Discussion}\label{sec:conclusion}

In the present paper, a computationally efficient constrained approach for clusterwise regression modeling was presented. Starting from the baseline idea of Seo and Lindsay (2010) and Seo and Kim (2012), we propose a new, computationally faster, data driven method to tune $c.$ Based on the simulation study and the two empirical applications, we have shown that the proposed method compares very well with the RGD method in terms of accuracy of parameter estimates and cluster recovery, doing from twice up to ten times faster than the RGD approach. 

In addition, we have demonstrated that the issue of unboundedness is not only an estimation problem, but seriously affects also the assessment of the number of components. We have implemented and deeply tested a formulation of the BIC, in the spirit of Fraley and Raftery (2007), using the (log) likelihood evaluated at the constrained solutions. To take into account the proportion of estimated scale entailed by the constrained estimator, we have also applied Cerioli et al (2017)'s recent proposal in our context, counting the number of free scales as the proportion (1-$c$) of unconstrained variances. In the simulation study and the empirical applications, we have shown that both approaches to compute the BIC based on the constrained estimator yield a sounder assessment of the number of components than standard unconstrained approaches. Cerioli et al (2017)'s correction seems to improve over the constrained BIC for the computationally more efficient approach and, in general, in relatively more complex modeling scenarios - i.e. larger number of (mixed--type) covariates (\emph{Auto--Mpg} data).

Having one tuning parameter to set ($k$) rather than two or more - as, for instance, in cross-validation schemes - limits the users' arbitrariness. In the real-data applications we observed that different values of $k$ might determine different conclusions on the chosen scale balance. In general, based on our results and having the cross-validated method as benchmark, larger values of $k$ (relative to the sample size) seems to be more favorable. 

In the simulation study, we found that selecting the number of components with a BIC based on the estimates of the homoscedastic normal (HomN) algorithm might work in some cases (smaller sample size and components with similar class sizes). Nevertheless, there are situations, like the one we analyzed in our second application, where also the BIC based on HomN overstates the number of components. Since neither of the two scenarios can be recognized \emph{a priori}, we suggest the use of BIC based on the constrained solutions to correctly assess the number of components.   

The equivariance property of our approach comes from the fact that the constraints are centered at a \emph{target} variance, which we re-estimate if the dependent variable is transformed. Having an equivariant method for clustering is crucial. The reason is not limited to requiring that the final clustering remains unaltered as one acts affine transformation on the variable of interest: more broadly, no matter how the data come in, affine equivariance means that there is no data transformation ensuring better results, since the method is unaffected by changes of scale in the response variable.

The approach from Cerioli et al (2017) that we have applied to clusterwise regression modeling is based on the consideration that by imposing constraints on the component variances, we are not estimating the full model scales, but only some fraction of them. Still, how this relates to the notion of effective degrees of freedom requires further research (see, for instance, Zou, Hastie, and Tibshirani, 2007).

\end{document}